\documentclass[aps,prl,twocolumn,superscriptaddress,groupedaddress]{revtex4}
\usepackage{graphicx}
\usepackage{dcolumn} 
\usepackage{bm}   
\usepackage{color}
\usepackage{ulem}
\usepackage{amssymb} 
\usepackage{natbib} 

\begin{document}
\setlength\parindent{5pt}
\widetext
\title{An Inelastic X-Ray investigation of the Ferroelectric Transition in SnTe }
\author{Christopher D. O'Neill}
\affiliation{School of Physics and Astronomy and CSEC, University of Edinburgh, Edinburgh, EH9 3JZ, UK}
\author{Dmitry A. Sokolov}
\affiliation{School of Physics and Astronomy and CSEC, University of Edinburgh, Edinburgh, EH9 3JZ, UK}
\affiliation{Max-Planck-Institut f\"{u}r Chemische Physik fester Stoffe, D-01187 Dresden, Germany}
\author{Andreas Hermann}
\affiliation{School of Physics and Astronomy and CSEC, University of Edinburgh, Edinburgh, EH9 3JZ, UK}
\author{Alexei Bossak}
\affiliation{ID28, ESRF, 71 avenue des Martyrs, 38000 Grenoble, France}
\author{Christopher Stock}
\affiliation{School of Physics and Astronomy and CSEC, University of Edinburgh, Edinburgh, EH9 3JZ, UK}
\author{Andrew D. Huxley}
\affiliation{School of Physics and Astronomy and CSEC, University of Edinburgh, Edinburgh, EH9 3JZ, UK}
\date{\today}
\begin{abstract}
\noindent  
\small

We report that the lowest energy transverse-optic phonon in metallic SnTe softens to near zero energy at the structural transition at $T_C=75 \text{~K}$ and importantly show that the energy of this mode below $T_C$ increases as the temperature decreases. Since the mode is a polar displacement this proves unambiguously that SnTe undergoes a ferroelectric displacement below $T_C$. Concentration gradients and imperfect stoichiometry in large crystals may explain why this was not seen in previous inelastic neutron scattering studies.  Despite SnTe being metallic we find that the ferroelectric transition is similar to that in ferroelectric insulators, unmodified by the presence of conduction electrons: we find that (i) the damping of the polar mode is dominated by coupling to acoustic phonons rather than electron-phonon coupling (ii) the transition is almost an ideal continuous transition (iii) comparison with density functional calculations identifies the importance of dipolar-dipolar screening for understanding this behaviour.

\end{abstract}
\maketitle

\indent SnTe was originally studied in the context of lattice vibrations in diatomic lattices \cite{Pawley}. There has been a recent resurgence of interest following its identification as a crystalline topological insulator, 
which is intimately related to its room temperature structure (the $fcc$ rocksalt structure) \cite{Hsieh, Tanaka, Li16}.   Stoichiometric SnTe is expected to be semiconducting with the minimum gap in the bandstructure of approximately 0.1 eV at the $L$ point in the Brillouin zone ~\cite{Cohen,Rabii,Littlewood_Arpes}.  Perfectly stoichiometric SnTe has, however, never been grown. Instead the crystals are always Te rich \cite{Brebrick, savage}, with the extra Te being accommodated in the lattice by Sn vacancies.  The vacancies lead to a high free carrier concentration of holes, $n_h$, that do not freeze out at low temperature.  
 Moreover, the material undergoes a phase transition to a rhombohedral structure upon cooling.  Such a structural transition would then strongly affect some of its topologically protected states \cite{Hsieh}, as has been reported in the Pb$_{1-x}$Sn$_{x}$Se class of crystalline topological insulators \cite{Okada}. The exact transition temperature, $T_c$, is dependent on the value of $n_h$ \cite {Kobayashi}. 

While the transition is predicted to be a displacive ferroelectric transition \cite{Littlewood_crystalstructure2, Littlewood_crystalstructure, Littlewood_coelastic, Rabe}, no ferroelectric response has previously been seen in bulk samples due to screening by the free carriers.  However a ferroelectric like response has been reported recently in ultrathin films \cite{Chang}. Whether this response co-exists with metallic conductivity is not clear.  A shift of much of the electronic density of states (eDOS) within 2 eV below $E_f$ was seen with ARPES \cite{Littlewood_Arpes}, suggesting an electronic mechanism. It is unclear whether the polarisation is the primary order parameter or a secondary order parameter as in some manganites \cite {Efremov}. For an insulator the divergence (or not) of the dielectric constant at $T_C$ resolves this issue. For metallic SnTe direct evidence establishing that polarisation is the primary order parameter is however missing.

 The non-stoichiometry means that SnTe is not insulating but a degenerate semiconductor (with a finite resistivity at low temperature).
Ferroelectricity and metallic conductivity are different macroscopic responses, with the former involving the separation of localized charges, while the latter is a property of delocalized electrons.  Since local charges are strongly screened by conduction electrons, the two properties might be considered to  be mutually exclusive.  Theoretically, they may however co-exist \cite{Anderson}, and such a coexistence was found in recent experimental studies of BaTiO$_{3-\delta}$ \cite{Kolod10} and LiOsO$_3$ \cite{Boothroyd}.  Co-existence is possible for electron concentrations up to a critical value at which the Thomas-Fermi screening length falls below a critical correlation length for ferroelectricity \cite{Kolod10}.

\begin{figure}[t]
\includegraphics[scale=0.22]{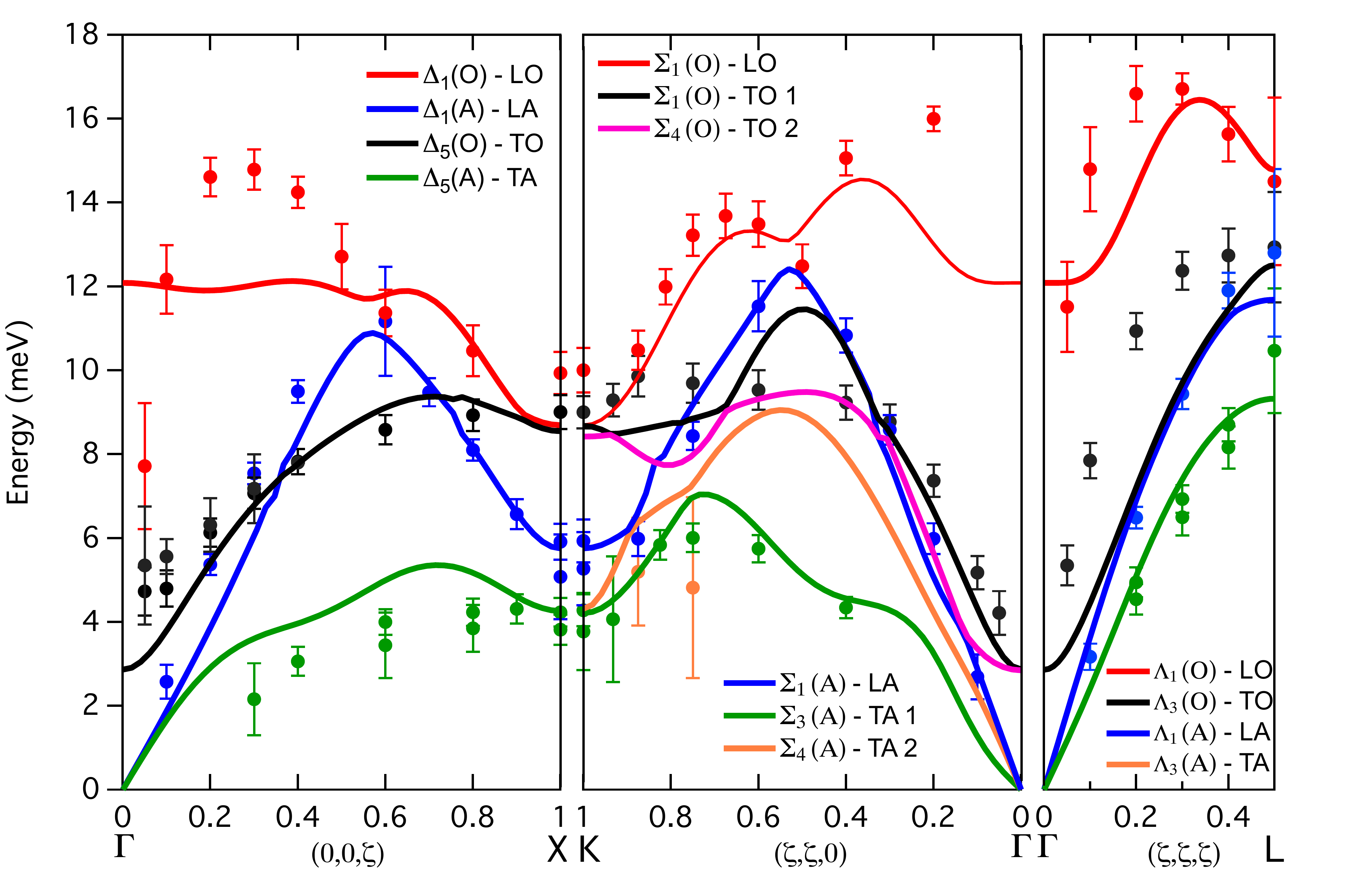}
\caption{\footnotesize The phonon dispersion for SnTe at $300 \textrm{ K}$ in the reduced Brillouin zone of the conventional cubic cell. The markers are measured experimental points from the phonon annihilation energy transfer while the lines are calculated phonon dispersion curves for $fcc$ SnTe at 300 K.  }
\end{figure} 
Generally, ferroelectric displacive transitions are expected to be accompanied by changes in the dynamics of the lattice with the energy of a  soft phonon mode at some high symmetry $\bf q$-points (either at the zone centre or edge) decreasing on cooling to a minimum at $T_c$, only reaching zero for a continuous transition and then rising \cite{Shirane, Cochran}. 
A typical example is the $1^{st}$ order structural transition from cubic to a ferroelectric tetragonal phase in PbTiO$_3$ \cite{Shirane_PbTi, Kempa, Tomeno}, which is driven by a soft transverse optic mode at the Brillouin zone centre. Empirically ferroelectric transitions are almost all  first order.  The 
reason for this is not clear although the long range nature of the dipole-dipole interactions is sometimes credited  \cite{Lines}. The coupling between the polarisation and strain further favours a first-order instability \cite{Chandra}. In contrast for SnTe we report that the phonon frequency approaches zero at the structural transition consistent with a continuous or very weakly first order transition.   \\

There are many examples of nearly continuous structural distortions in metals such as the cubic to tetragonal transition in V$_3$Si and Nb$_3$Si \cite{Testardi} driven by a soft acoustic mode towards the zone edge.  However these martensitic transitions are not ferroelectric. Our identification that a zone centre transverse optic phonon softens and recovers going through $T_c$ in SnTe indicates its transition is a ferroelectric transition like that in PbTiO$_3$.

\begin{figure}[t]
\includegraphics[scale=0.31]{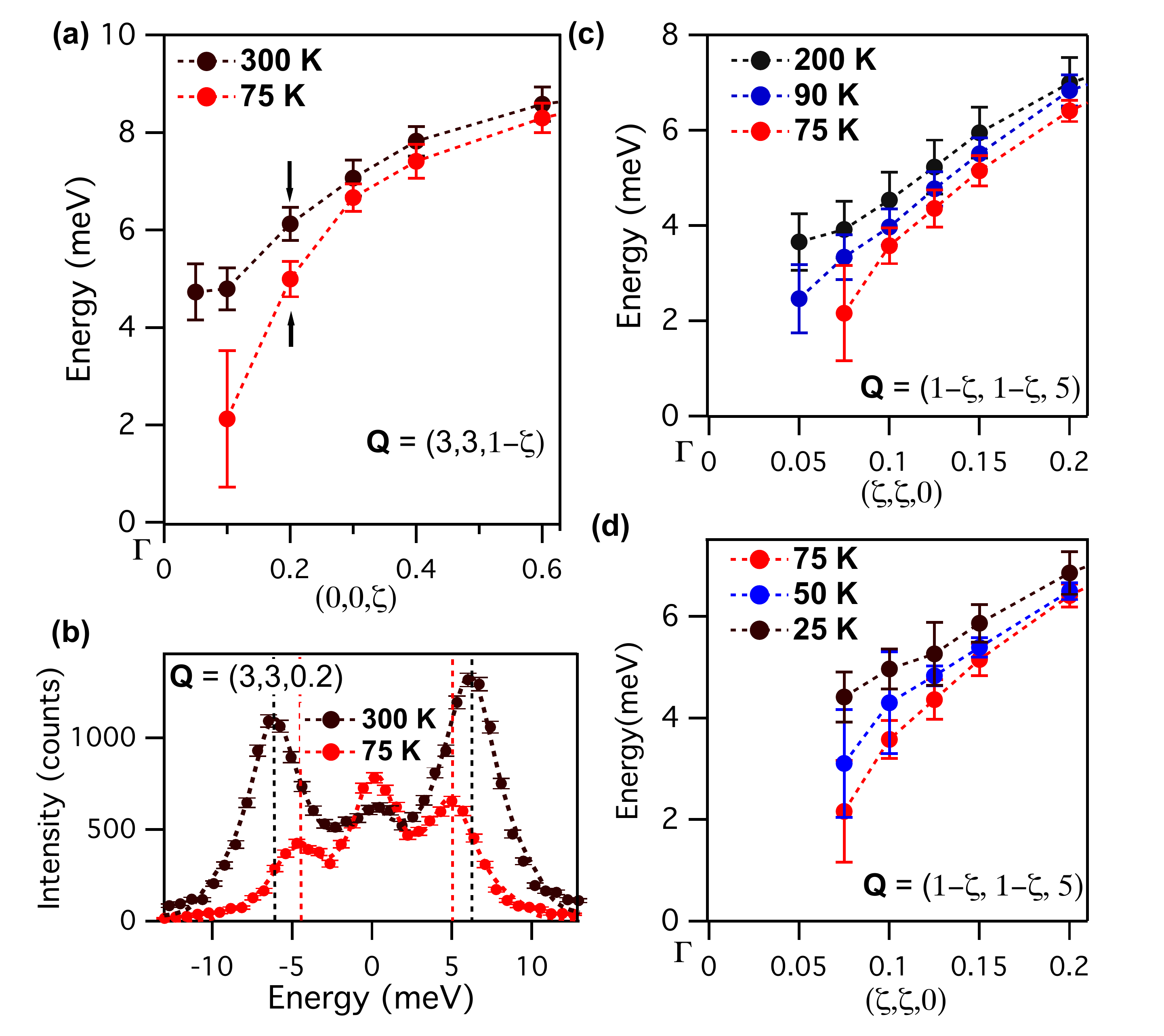}
\caption{\footnotesize {\bf (a)} The transverse optic phonon energy measured for a total momentum transfer ${\bf Q} = (3,3,1) \rightarrow (3,3,0)$ at 300 K and 75 K.    The figure shows the suppression of the phonon energy at the ferroelectric transition temperature $T_c=75 K$.  {\bf (b)} The measured intensity (data points) vs energy for the TO phonon at ${\bf Q} = (3, 3, 0.2)$ along with fitted lines described in the text. The peak positions in these spectra (vertical lines) give the plotted points indicated with black arrows in (a).  {\bf (c)}  The transverse optic phonon measured for ${\bf Q} = (1,1,5) \rightarrow (0,0,5)$ for a range of temperatures above $T_c $ and {\bf (d)} below $T_c$ where the phonon energy increases again below $T_c$. }
\end{figure} 

\indent Previous investigations of the phonon dispersion curves in SnTe carried out by Pawley $et\; al.$ \cite{Pawley} {and more recently Li $et\; al.$ \cite{Li14}} using inelastic neutron scattering found significant softening of the transverse optic phonon energy towards the Brillouin zone centre ($\Gamma$ point). Similar softening has been reported in the binary material Pb$_{1-x}$Sn$_x$Te \cite{Dolling_Pb}.  Although the softening was strongly temperature dependent, phonon energies never softened close to zero even at the lowest measured temperatures.  Therefore these authors suggested SnTe and  Pb$_{1-x}$Sn$_x$Te  approached ferroelectric transitions without ever passing through them. However indications of a structural distortion have been reported in other experiments such as powder X-ray diffraction where peak splitting was seen consistent with the cubic phase being distorted along (1,1,1) directions to a rhombohedral shear angle of  $\delta \alpha \sim 0.115^\circ$  \cite{Muldawer}.  Other signatures of the transition are a  change of Bragg reflection intensity in neutron scattering \cite{ Iizumi} consistent with a relative shift of the two $fcc$ sublattices of $\tau \sim 0.007$ ($\sim 4$ pm), and changes in the Raman spectrum \cite{Brillson} and heat capacity \cite{Hatta}.  A cusp-like anomaly in the electrical resistivity at $T_c$  \cite{Koba75, Grassie} has been attributed to the presence of a soft phonon mode.  The samples previously studied with inelastic neutron scattering most likely had a value of $n_h$ that was sufficiently high for $T_c$ to have been suppressed.   A clear proof that the observed structural transition is a ferroelectric displacement is then lacking.   To 
determine whether SnTe is indeed ferroelectric, we performed inelastic x-ray measurements on a small, close to stoichiometric, single crystal. \\
\indent The single crystal was grown from equal molar weights of high purity elements Sn (99.9999$\%$) and Te (99.9999 $\% $) wrapped in Mo foil (99.95$\%$), sealed in an evacuated silica ampoule and heated to $850 ^\circ\textrm{C}$, well above the melting temperature \cite{Brebrick}.  The mixture was slowly cooled to $760 ^\circ\textrm{C}$ and the ampoule then quenched into water at room temperature.  The structural transition at $T_c=$75~K was identified in the extracted crystal from a clear change in the slope of the electrical resistivity (see supplementary material).  A value of $n_h = 3.23 \pm 0.05 \times 10^{20}\textrm{cm}^{-3}$ was deduced from a measurement of the Hall resistivity at 2~K (see supplementary material).   
The value of $T_c$ = 75 K is in good agreement with previous measurements for similar $n_h$ \cite{Kobayashi}.  \\
\indent Following an initial examination with diffuse scattering (supplementary material), the single crystal was studied by inelastic X-ray scattering with the ID28 instrument at the ESRF (Grenoble). A backscattered silicon (9, 9, 9) monochromator was used. The scattered photons were analysed with 9 single crystalline spherical silicon analysers mounted in a pseudo Rowland circle geometry.  Energy scans were performed at constant $\bf Q$-vector by varying the temperature of the monochromator covering the range -19.82meV to 19.82meV in 0.68 meV steps centred at the analyser energy of 17.794 keV. The  penetration depth of the incident X-rays was $45 \mu$m, ensuring reported results are bulk properties. The energy resolution had a Lorentzian lineshape with a 3meV FWHM.   
Measurements were made over different zones to distinguish the different phonons.   The experimental spectra are well described by the instrument resolution convolved with antisymmetrised pairs of Lorentzians for each mode, respecting detailed balance \cite{Stock}, plus a narrow fixed width Lorentzian at the elastic position (see FIG 2 (b) for an example).  \\

\indent Electronic structure calculations were performed for the stoichiometric compound,  using scalar-relativistic density functional theory (DFT) as implemented in the {\sc VASP} package, \cite{VASP.Kresse1996,PAW.Joubert1999,GGA.PBE1996}, a plane wave basis with cutoff energy $E_c=200$~eV  and k-point sampling with a linear density of $30/\text{\AA}^{-1}$  (grid size $(8,8,8)$). The energy difference between the cubic and rhombohedral structures converged to within 0.05~meV/formula. We found the ground state energy for the rhombohedral structure to be about 0.2~meV/formula lower than for the cubic structure, with an optimum rhombohedral angular distortion of $\delta\alpha\approx 0.2^{\circ}$ and a ferroelectric displacement of $\tau\approx 0.009$. Ground state phonon dispersion calculations used the finite displacement method \cite{Alfe2009} in 128-atom supercells and confirmed the dynamical stability of the rhombohedral phase and the instability of the cubic phase (supplementary material). The self-consistent ab-initio lattice dynamics method (SCAILD) was then used to obtain phonon dispersion curves at finite temperatures for the cubic phase \cite{SCAILD.Souvatzis2008}. The calculated phonon dispersion for the two structures (rhombohedral at T$=0$ and cubic at T$=300~$K) are very similar over most of the Brillouin zone. The main differences are close to the zone centre where there is a sharp reduction of the energy of the LO ($\Delta_1(O)$) mode for the rhombohedral structure and a more modest reduction of the TO ($\Delta_5(O) \rightarrow \Delta_2(O)+\Delta_5(O))$ mode (figure 1 and supplementary material). The former may be a consequence of modulations of the polarisation present in the rhombohedral structure more effectively screening the LO phonon.  \\

\begin{figure}[t]
\includegraphics[scale=0.3]{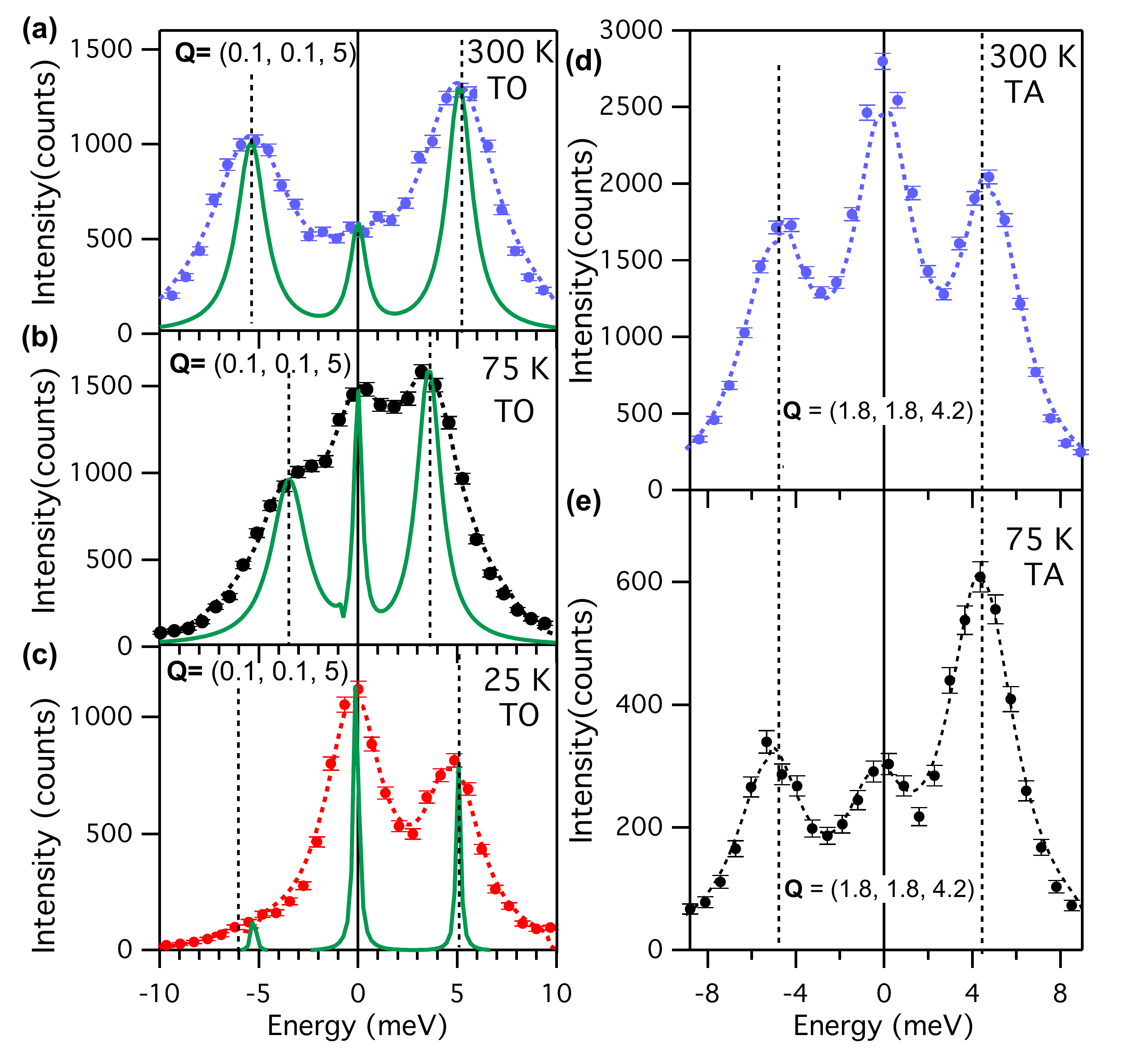}
\caption{\footnotesize{\bf (a)-(c)}  Measured intensity vs energy transfer spectra at various temperatures shown as markers along with calculated fits (dashed lines) at ${\bf Q} = (0.1, 0.1, 5)$ for the transverse optic phonon propagating in the [1,1,0] direction.  The solid green lines, are the  peaks from the calculated fits unconvoluted from the instrument resolution.   {\bf (d)-(e)} Measured intensity vs energy transfer and fits at 300 K and 75 K respectively at ${\bf Q} = (1.8, 1.8, 4.2) $ for the transverse acoustic phonon propagating in the [1,1,1] direction. }
\end{figure} 

\indent In Figure 1 the cubic phase calculation is compared with the measured phonon dispersion curves at 300 K (momentum vectors are given with respect to the conventional cubic unit cell all throughout the paper). No experimental points for the $\Sigma_4(O)$ branch could be accurately determined. Our measurements agree with previous inelastic neutron scattering measurements above $T_c$ \cite{Cowley}.   The sharp decrease in the LO phonon energy towards $\Gamma$ and dip along the (1,1,0) direction in the measurements relative to the calculation have previously been attributed to the macroscopic electric field associated with the phonon coupling to free charge carriers \cite{Cowley_LO, Cowley}.   {The LO phonon dips over a range of $\zeta < 0.2~a^*$ around $\Gamma$ (figure 1). The reciprocal-space width over which the the TO phonon shows a temperature dependence ($\zeta > 0.2~a^*$, figure 2 a) is larger than this, supporting the hypothesis that the Thomas Fermi-screening length exceeds the ferroelectric correlation length.  However, the LO dispersion measured with neutrons by Pawley \cite{Pawley} is indistinguishable from our measurements, although a larger range for the dip, reflecting a shorter Thomas-Fermi screening length, might have been expected from the absence of ferroelectric order in their crystal. This is reconciled by noting that the polarisation may provide the dominant screening mechanism, explaining the dip in the LO mode \cite{Cowley}.}
\\
\indent Phonon dispersion curves at $T_c$ = 75 K are shown in the supplementary material. There is no measurable change of the experimentally determined elastic constants between 300 K and 75 K consistent with previous measurements \cite{Beattie, Littlewood_coelastic}.
The only observable temperature dependence  is for the TO mode approaching $\Gamma$ and we focus on this in the following. 
\\  
\indent The measured energies of the transverse optic (TO) phonon at 300 K and at $T_c$ = 75 K 
are shown in Figure 2 (a).   A softening of the phonon energy towards the Brillouin zone centre, $\Gamma$, at low temperature is clearly seen.  Measured spectra at ${\bf Q} = (3,3,0.2) $ are shown in Figure 2 (b) at 300 K and 75 K;  the 3 peaks correspond to phonon creation (negative energy),  phonon annihilation (positive energy) and a small elastic contribution at zero energy.  The resolution function makes very low energy phonons difficult to distinguish from the elastic line close to the zone centre where the Bragg condition is met. Thus, in Fig 2 (a) the 75 K data is missing a point at ${\bf Q} = (3, 3, 0.95)$.  Further temperature points were measured for the TO phonon along the (1,1,0) direction. This confirms the softening on cooling to $T_c$ (Figure 2 (c)), while $\textrm{Figure 2 (d)}$ shows that on further cooling below $T_c$ to 50 K and then to 25 K the phonon energy recovers.   Individual scans at ${\bf Q} = (0.1,0.1,5)$ are shown in Figures 3 (a-c),  along with calculated fits unconvoluted from the instrument resolution to judge their reliability. For comparison the measured spectra for the transverse acoustic (TA) phonon at 300 K and $T_c$ are shown in Figures 3 (d) and (e) at  ${\bf Q} = (1.8, 1.8, 4.2) $, the closest point to $\Gamma$ that could be measured accurately, showing that this mode does not change with temperature.
\\
\indent The values of phonon energies at $\Gamma$ were determined by linearly extrapolating the phonon energy squared $E({\bf q})^2$ plotted against ${\bf q}^2$ to ${\bf q}^2 \rightarrow 0$ (supplementary material), where {\bf q} is the reduced momentum.  Figure 4 (a)  shows that $(E_{\text{TO}}(\Gamma))^2$ decreases linearly with temperature to almost zero at $T_c$ and increases again above $T_c$. In the Landau theory of ferroelectrics the ratio of the slopes $d((E_{\text{TO}}(\Gamma))^2)/dT$  below $T_c$ to above $T_c$ should be -2. The electrical  resistivity was previously found to be adequately described with a simple model calculation based on this ratio for a temperature range $-0.35<(T-T_c)/T_c<1.4$  \cite{Koba75}. The solid line in the figure is a fit assuming the Landau theory ratio and including a saturation of $E({\bf q})^2$ towards a fixed value  at high temperature reflecting the much larger temperature range $-0.35<(T-T_c)/T_c < 4$ spanned by our measurements.
\\
\indent The continuous `$2^{nd}$ order' nature of the transition,
previously suggested by heat capacity measurements \cite{Hatta}, is reflected in Figure 4 (a) by the nearly complete softening to zero energy of $E_{\text{TO}}(\Gamma)$ at $T_c$.  Such a complete softening of the phonon energy is highly unusual for ferroelectric transitions in insulators.  \\
\begin{figure}[t]
\includegraphics[scale=0.29]{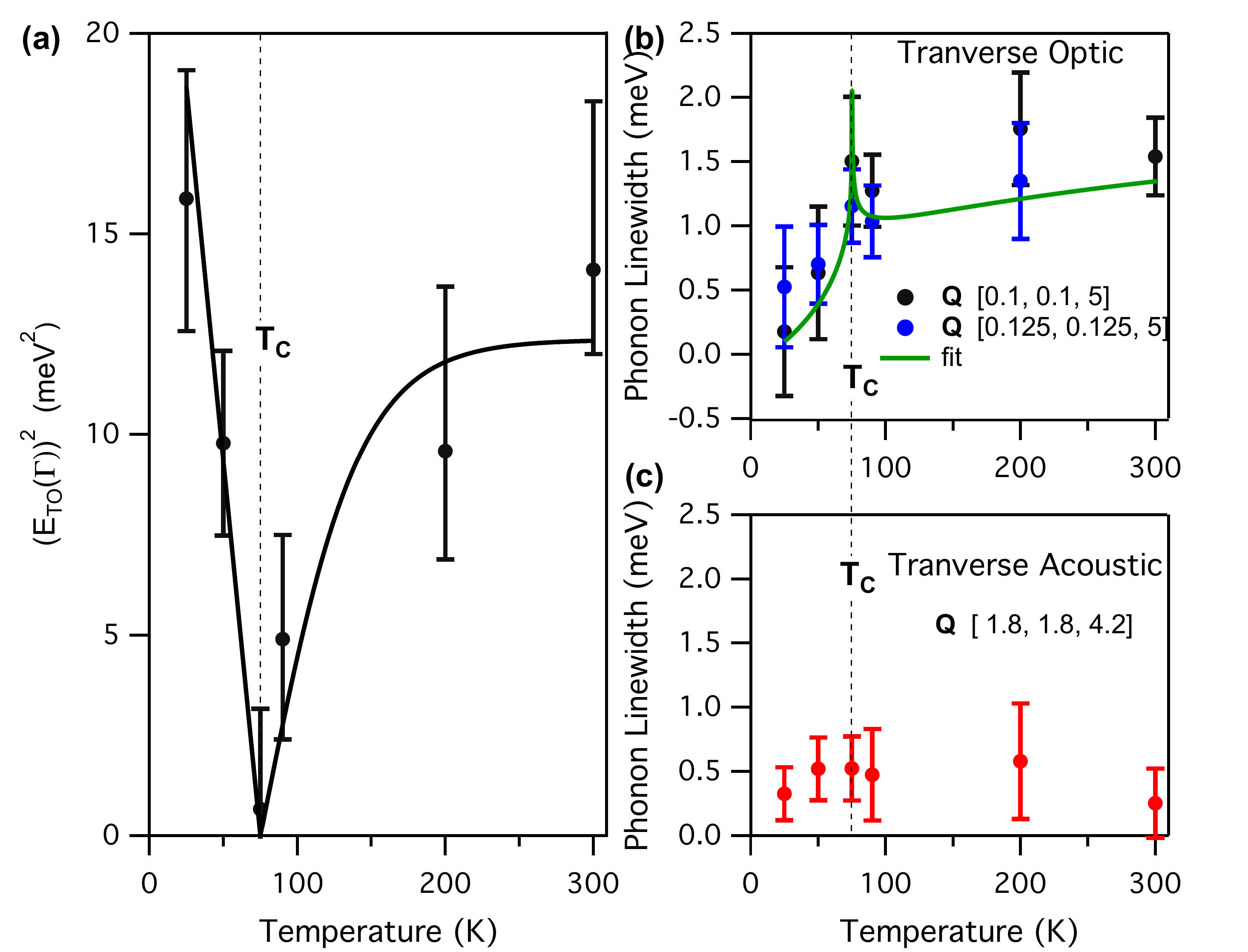}
\caption{\footnotesize (a) The extrapolated energy squared, $(E_{\text{TO}}(\Gamma))^2$, of the zone-centre TO phonon plotted against temperature.  The solid line is explained in the text. The phonon is seen to soften to zero energy within the experimental resolution at $T_c$.  (b) The variation of the TO phonon line-width with temperature at ${\bf Q} = (0.1, 0.1, 5)$ and ${\bf Q} = (0.125, 0.125, 5)$. The line-width is seen to be large above $T_c$, but is suppressed below $T_c$.    The calculation for a simple model of anharmonicity described in the text is shown by the solid line. (c) The temperature dependence of the TA phonon line-width at  ${\bf Q} = (1.8, 1.8, 4.2)$. The TA width is smaller than the TO width and has no  significant temperature dependence.  }
\end{figure} 
\indent The coupling of LO and TO optic modes has been argued to be responsible for a large phonon non-harmonicity in PbTe and SnTe \cite{Dela11, Li14} that may explain the high figures of merit of these materials for thermoelectric applications. We found no temperature dependence or abnormal damping of the LO phonon to support this.
\\
\indent We now discuss the line-width of the TO phonon. The line-width as a function of $\bf q$ increases sharply towards $\Gamma$ above $T_c$ but is small and constant in $\bf q$  below $T_c$ (supplementary material).  Figure 4(b) shows the line-width as a function of temperature at ${\bf Q} = (0.1,0.1,5) $ and ${\bf Q} = (0.125,0.125,5)$. The TO line-width is enhanced at the zone centre above $T_c$ and this enhancement is suppressed below $T_c$. Figure 4(c) shows the line-width of the TA phonon at ${\bf Q} = (1.8,1.8,4.2)$ for comparison which is much smaller and shows no significant temperature dependence.
\\   

 We find that a very simple model for anharmonicity based on the phonon interaction $TO(0)+TA({\bf q}) \leftrightarrow TO({\bf q})$ or $TO(0)+LA({\bf q}) \leftrightarrow TO({\bf q})$ \cite{Dvorak} can explain our measurements. The calculated curve in figure 4 (b) is determined from $E_{\text{TO}}(\Gamma)$ taken from Figure 4(a) (solid line) and a fixed intercept with an acoustic phonon branch at 4 meV (see supplementary material). There is only one further parameter in the calculation that fixes the overall amplitude of the damping. This parameter is proportional to the ratio $Q$ of the strain $\epsilon$ induced by the displacement to the squared ferroelectric displacement $\tau^2$. The value required to fit the data gives $Q=\epsilon/\tau^2 = 17$. This value of $Q$ agrees well with an estimate from our DFT calculation $Q_{\text{DFT}} = 13.4$ which also reproduces the observed magnitudes of $\alpha$ and $\tau$. This strongly suggest that phonon anharmonicity from coupling to acoustic phonons indeed explains the measured TO phonon line-width.

 The electron TO-phonon interaction has been considered to drive the displacement transition \cite{Kobayashi}.  Although this interaction and its modification with doping may play a major role in determining the TO phonon energy our analysis shows that it contributes only indirectly to the phonon linewidth via the phonon energy and the strain-displacement coupling.  A direct electron-phonon contribution to the line width from the conduction electrons is not needed, however can not be ruled out. 
We also note that there is no significant change in the Hall resistivity between 150 K and 2 K (supplementary material), ruling out changes in the number of conduction electrons as the cause of the drop in the line width below $T_c$.

In summary, we have shown for the first time that the structural transition in a metallic sample of SnTe is a ferroelectric displacement transition with the TO phonon hardening below $T_c$. Such a recovery of the phonon energy below $T_c$ has never been reported before, despite being highly sought after, and confirms that polarisation is the primary order parameter. We found that the damping of the TO phonons close to the zone centre can be explained by a conventional coupling of displacement and strain. Such a coupling acts to make the transition first-order. The continuous (or only weakly first order) nature of the transition then requires that the dipole mode-mode coupling term in a Landau expansion of the free energy (the $4^{th}$ order term in powers of the polarisation) is large in the absence of strain. The marked difference of the measured LO phonon energy from that calculated with DFT suggests that a strong many-body dipole-dipole screening may be present in the parent insulating material. Identical LO phonon energies across samples where the structural transition is present or absent also suggest that $T_c$ is not suppressed by screening of the dipole-dipole interaction through added charge carriers.  Another mechanism for the suppression is provided in ref. \cite{Kobayashi} where it is suggested that it is due to the removal of valence electrons rather than screening from conduction electrons. Our results support such a diminished role for the conduction electrons and indicate that their contribution to characteristics such as the soft-mode phonon line-width is minor.


{\footnotesize \bf{Acknowledgements}}
{\footnotesize Support from the Royal Society RG-150247 (A.H.), Engineering and Physical Sciences Research Council EP/L0151101/ and EP/J00099X (CON) and EP/I031014 (ADH) and Carnegie 
Trust for the Universities of Scotland (C.S.) is acknowledged.}

\widetext
\begin{center}
\textbf{\large Supplemental Material; An Inelastic X-Ray investigation of the Ferroelectric Transition in SnTe }
\end{center}

\setcounter{equation}{0}
\setcounter{figure}{0}
\setcounter{table}{0}
\setcounter{page}{1}

\renewcommand{\theequation}{S\arabic{equation}}
\renewcommand{\thefigure}{S\arabic{figure}}
\renewcommand{\bibnumfmt}[1]{[S#1]}
\renewcommand{\citenumfont}[1]{S#1}

\noindent{\bf \underline{Transport Measurements}}

The resistivity and Hall resistivity of the SnTe single crystal were measured with a 100 $\mu$A ac (37 Hz) current applied along the [1,0,0] direction in a standard 6 terminal configuration. $T_c$ is taken as the point of a clear break in the temperature derivative of the resistivity at 75 K (supplementary figure 1 (b)).  The Hall signal  (supplementary figure 1 (c)) has been anti-symmetrised for positive and negative applied field to correct for the small contact misalignment.  The initial slope of the Hall resistivity against field determines the carrier concentration $n_h = 3.23 \pm 0.12 \times 10^{20}\textrm{cm}^{-3}$.

\begin{figure}[h]
\includegraphics[scale=0.29]{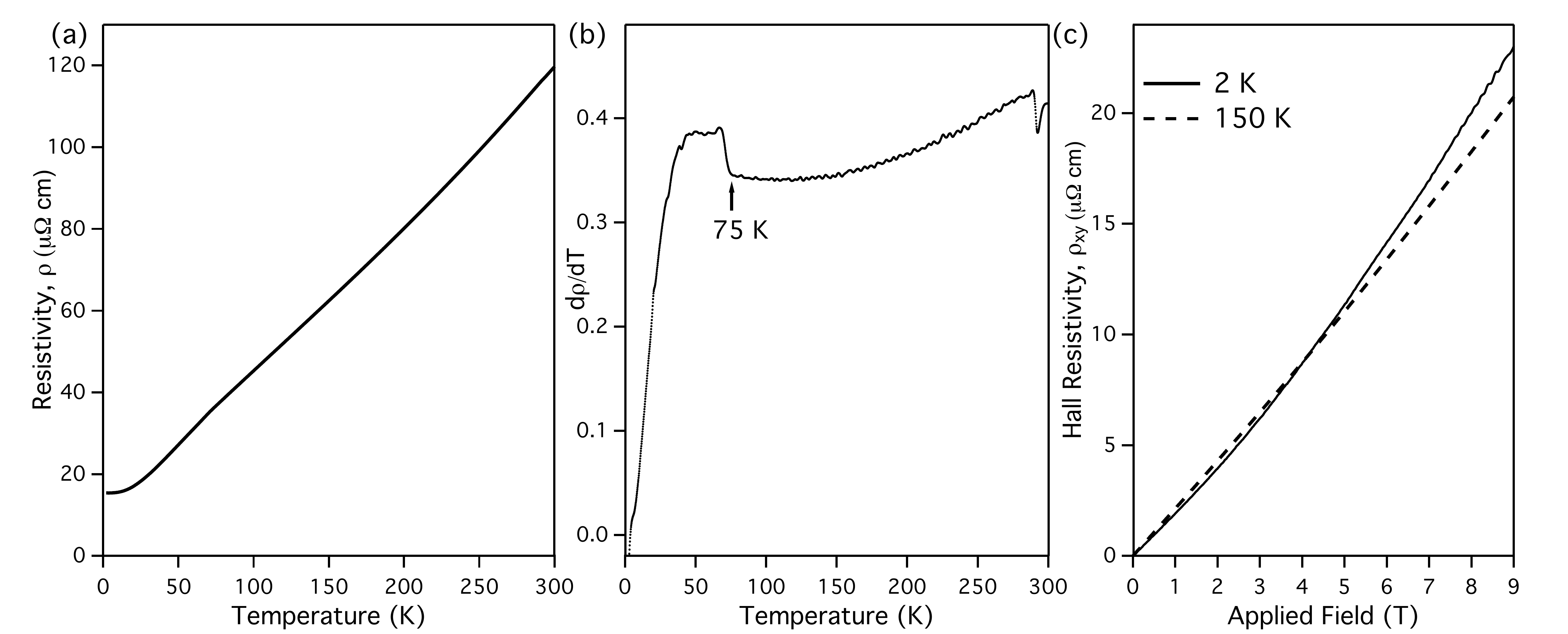}
\caption{ \label{fig}{\bf (a)} The electrical resistivity of the SnTe single crystal investigated by inelastic x-ray scattering,  {\bf (b)} the temperature derivative of the resistivity, and {\bf (c)} the Hall resistivity measured at 2 K and 150 K (dashed line).}
\end{figure}

\noindent{\bf \underline{Thermal Diffuse Scattering}}

  The crystal was polished following the transport measurements.  Prior to the inelastic scattering measurements thermal diffuse X-ray scattering was measured at 300 K, with the ID23 instrument at the ESRF. Two diffuse scattering intensity patterns measured in a transmission mode with an image plate detector are shown in supplementary figures 2 (a) and (b).  The incident wavelength was selected such that no Bragg condition is satisfied over the entire area of detection.  The scattering is therefore inelastic and the bright spots in the pictures correspond to scattering close to reciprocal lattice points where the acoustic phonon population is high \cite{Holt}.  Optic phonon modes also contribute to the measured intensity.  
These measurements show that the phonons can be easily detected with x-rays and confirm the long range order and high quality of the crystal. 

\begin{figure}[h]
{\bf (a)}\includegraphics[scale=0.065]{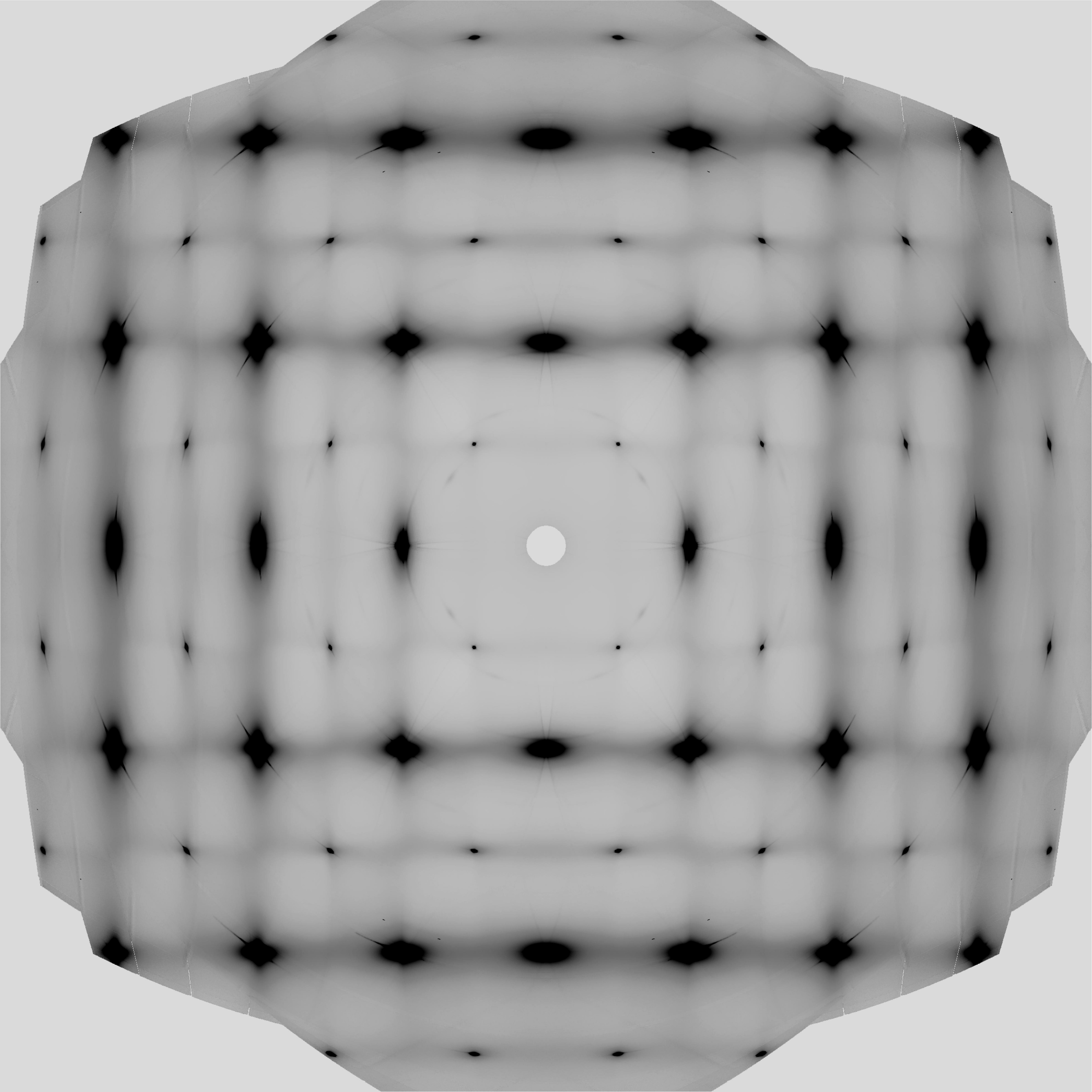}
{\bf (b)}\includegraphics[scale=0.065]{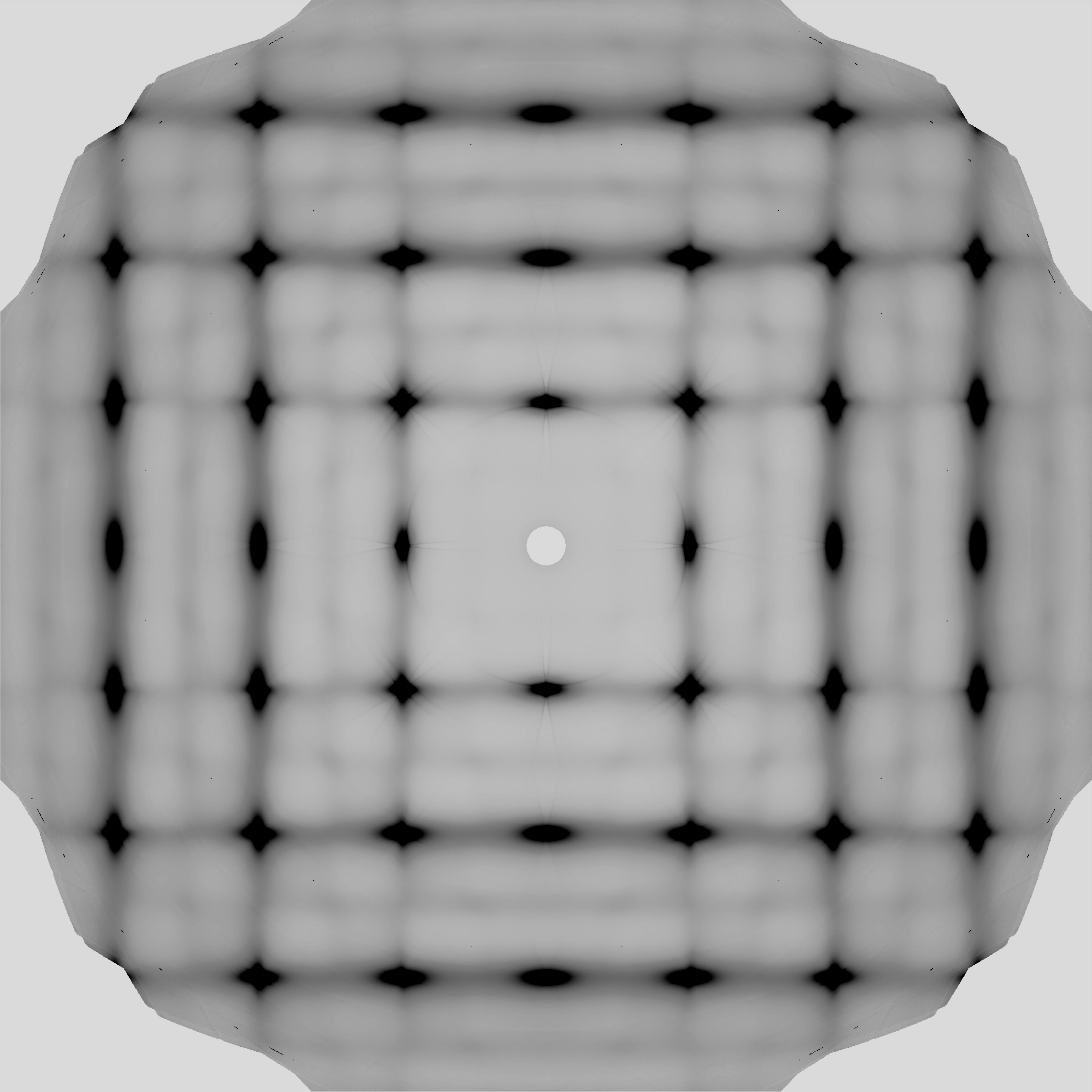}
\caption{ \label{fig}Thermal diffuse scattering patterns at 300 K of the single crystal of  SnTe for an incident beam along the [1, 1, 1] direction {\bf (a)} and the  [1, $\overline{1}$, 0] direction {\bf (b)}.    }
\end{figure}

\vspace{10mm}
\noindent{\bf \underline{Measured phonon dispersion at $T_c$ and calculated dispersion for the rhombohedral phase at 0 K}}

The complete phonon dispersion measured at $T_c$ is shown in supplementary figure 3. Also shown in this figure is the dispersion calculated at 0 K in the rhombohedral ground state (the calculation parameters are described in the main text). This is compared with the experimental data at 300 K and calculations for the cubic phase at 300 K (figure 1) in the main text. Additionally a small lifting of the degeneracy of the transverse modes along the [0,0,1] direction is apparent in the rhombohedral phase.  

\begin{figure}[h]
\includegraphics[scale=0.3]{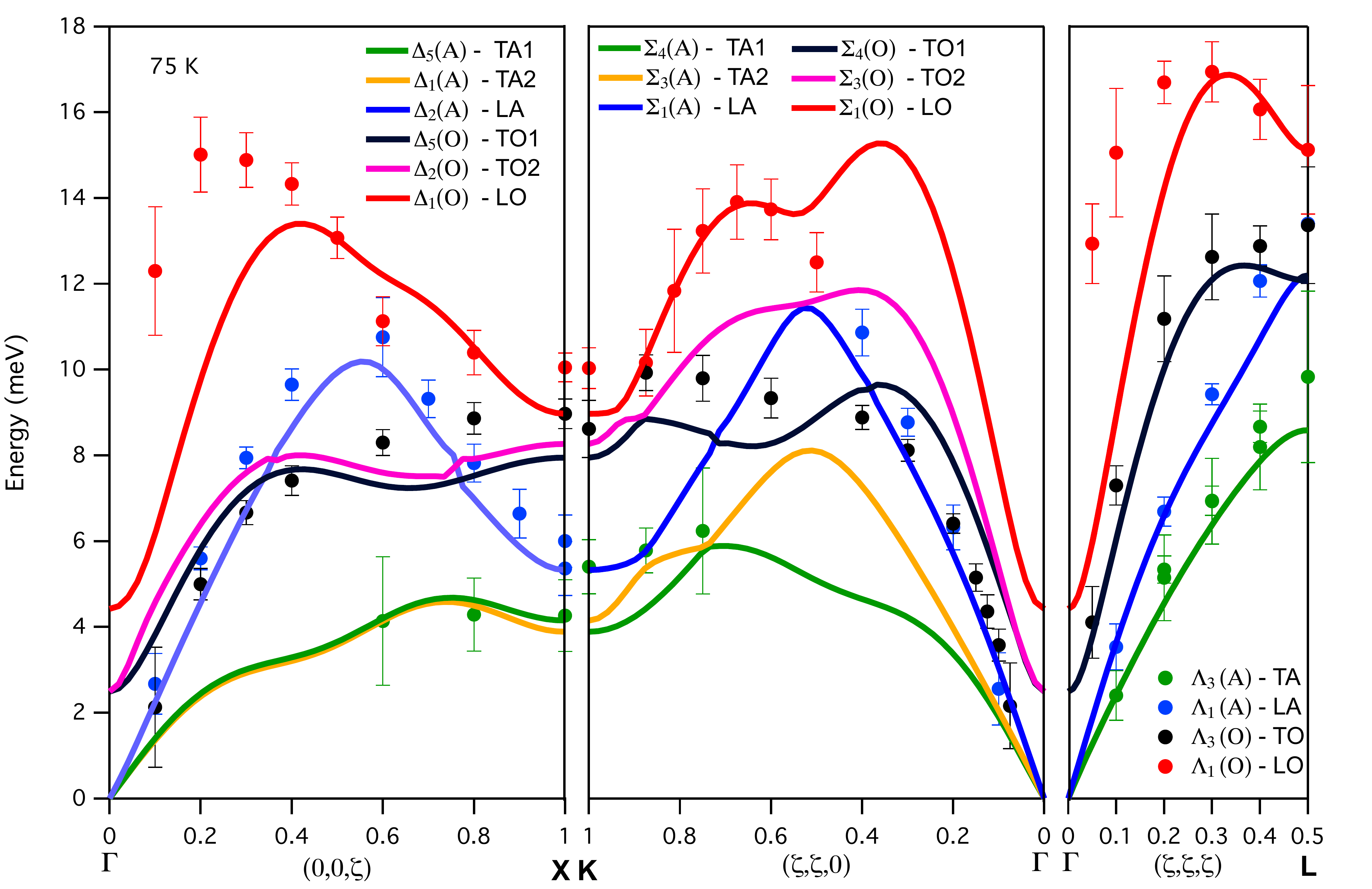}
\caption{ \label{fig}  Measured phonon dispersion curves along all 3 high symmetry directions for SnTe at $T_c$ = 75 K (points) and ground state 0 K calculation for the rhombohedral phase (lines).   }
\end{figure}

\noindent{\bf \underline{Calculated phonon dispersion for the cubic phase at 0 K}}

\indent {The phonon dispersion obtained for the cubic phase in the ground state is shown in supplementary figure 4. It clearly shows the degenerate dynamical instability at the $\Gamma$ point that corresponds to the ferroelectric distortion.}

\begin{figure}[h]
\includegraphics[scale=0.3]{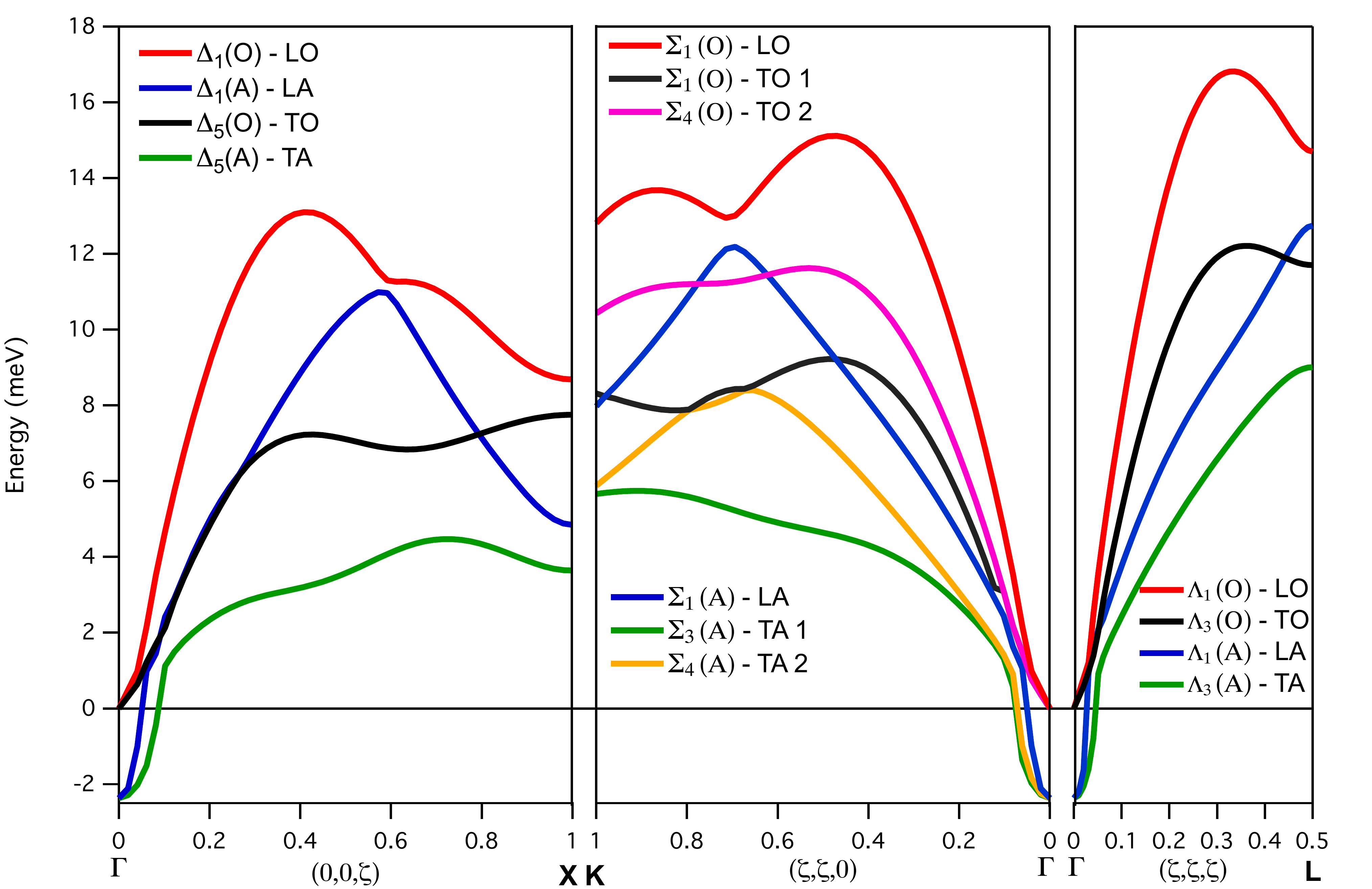}
\caption{ \label{fig:disp-cub}{The phonon dispersion in cubic SnTe, as obtained from finite-displacement DFT calculations.}}
\end{figure}

\vspace{3mm}
\noindent{\bf\underline{Determination of the TO-phonon energy at the zone centre}}

  To determine the TO phonon energy at the Brillouin zone centre $\Gamma$,  the measured phonon energy squared was plotted against the squared ${\bf q}$ vector and lines of best fit calculated and extrapolated to $\Gamma$, as discussed in the main text. The gradient of the lines was constrained to be independent of temperature at and above $T_c$ (supplementary figure 5 (a)). The gradient below $T_c$ was similarly constrained to be independent of temperature (supplementary figure 5 (b)). The slope for the low temperature data is clearly less than for the high temperature data. The point of interest is the low $\bf q$ intercept which is relatively insensitive to the extrapolation details and is plotted in figure 4 (a).  

  \begin{figure}[h]
\includegraphics[scale=0.29]{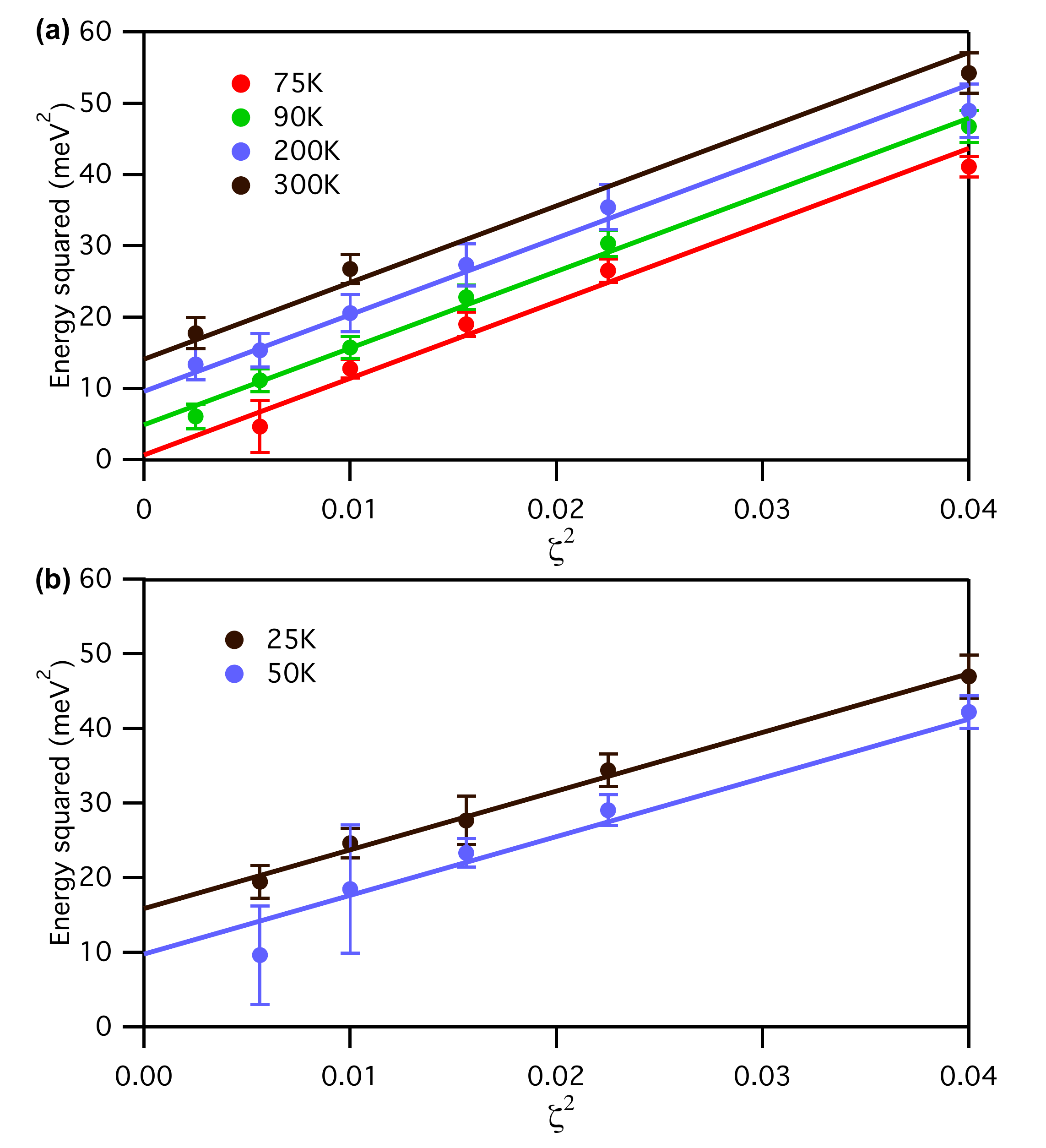}
\caption{ \label{fig}{\bf (a)} The phonon energy squared vs $\zeta^2$ where ${\bf q}=(\zeta,\zeta,0)$ at temperatures  $T_c$ and above, shown as markers, with the best fit linear lines constrained to have the same slope and  {\bf (b)} the data and fits below $T_c$.}
\end{figure}

\vspace{3mm}  
\noindent{\bf\underline{Linewidth as a function of q}}

The change in phonon linewidth as a function of $\bf q$ is shown in supplementary figure 6 for 300 K, $T_c$ = 75 K and below $T_c \textrm{ at 25 K}$ along the [110] direction.  

\begin{figure}[h]
\includegraphics[scale=0.35]{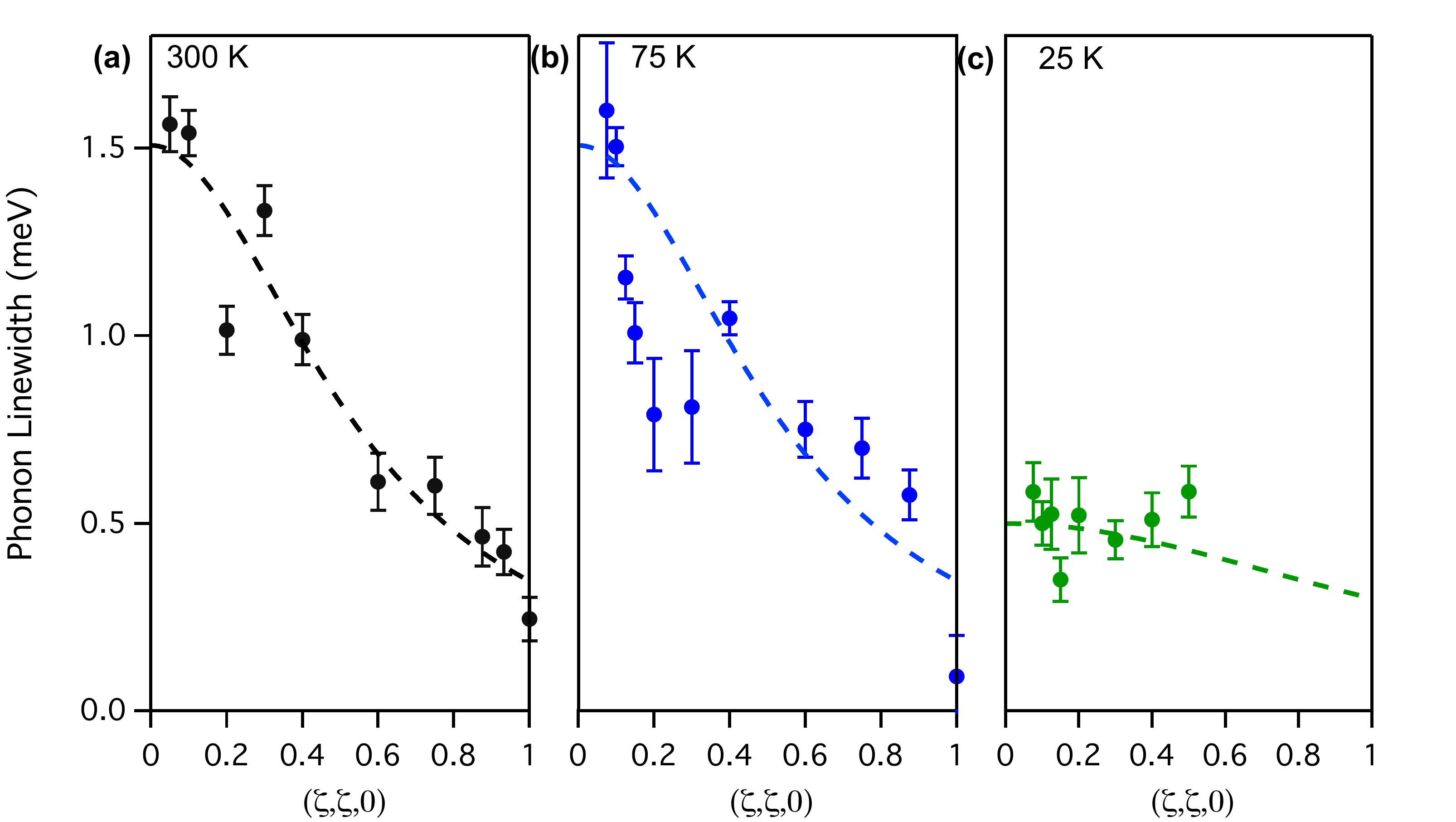}
\caption{ \label{fig}  Phonon linewidths as a function of $\bf q$ $\bf (a)$ above $T_c$, $\bf (b)$ at $T_c$  and $\bf (c)$ well below $T_c$.  The dashed lines are a guide to the eye.}
\end{figure}
\newpage
\vspace{3mm}
\noindent{\bf \underline{Phonon damping}}

In this section we consider the linewidth $\Delta \omega_{\text{TO}}$ for the zone centre TO phonon due to coupling with acoustic phonons.  

Momentum and energy conservation for the process $\omega_{\text{TO}}(0)+\omega_{A}({\bf q}) \leftrightarrow \omega_{\text{TO}}({\bf q})$ are satisfied on surfaces where the TO-phonon energy relative to its value at the zone centre  ($\omega_{\text{TO}}({\bf q})-\omega_{\text{TO}}(0)$) intercept an acoustic phonon energy $\omega_A({\bf q})$. Comparing the measured TO phonon at 75~K  (supplementary figure 3) shifted down by $\omega_{\text{TO}}(0) \approx 0.3 meV$ with the DFT-calculated acoustic phonon dispersions suggests there are potential crossing points with the LA phonon branch along $[\zeta,\zeta,0]$ and $[\zeta,0,0]$ close to 4 meV. At 300 K the dispersion of the TO mode shifted to zero energy (figure 1) is too weak to cut the LA branch, except possibly along $[\zeta,\zeta,\zeta]$, but now cuts the TA mode at similar energies to 4 meV. To keep the calculation as simple as possible we do not distinguish between the LA and TA phonons and consider that energy and momentum conservation occur at a fixed ${\bf q}={\bf q}_c$ and energy $\omega_A \sim 4 \textrm{meV}$ with the constant values of $d\omega/dk$ at the crossing points.
With this simplification the line width can be expressed as \cite{Dvorak1}

$$ \Delta \omega_{\text{TO}} = K \frac{n[\beta \omega_A]-n[\beta (\omega_{\text{TO}}(0)+\omega_A)]}{\omega_{\text{TO}}(0) (\omega_{\text{TO}}(0)+\omega_A)} $$
with $n[x] = (e^{x}-1)^{-1}$ and $\beta = 1/ k_B T$.

This formula was used to draw the solid line in figure 4b of the main text. It is instructive to proceed with an estimate of the damping based on the above simple model for which $K$ is 
$$K = \frac{16}{27 \pi} \frac{Q^2 c^2}{\rho^3 a^7} \frac{\hbar ({\bf q}_c a)^3}{v_s (v_s-v_{\text{TO}})}.$$
$v_s$ and $v_{\text{TO}}$ are the velocities of the acoustic phonon and TO phonon at the crossing point ${\bf q}_c$.  The LA phonon has a measured velocity $v_L = 3200 \textrm{~ms}^{-1}$ along $[\zeta,\zeta,0]$ which agrees well with ultrasound data and DFT \cite{Miller}. From the DFT calculation the energies of the TA phonon modes merge approaching $\Gamma$ along [110] consistent with an isotropic elastic response as assumed theoretically and a transverse sound velocity $\sim (2/3) v_L$. In contrast the elastic constants deduced from ultrasound measurements suggest that the TA1 mode would have a considerably lower velocity than TA2 along [110] \cite{Miller} and that the elastic response is far from isotropic, although the primary data supporting these conclusions was not given.  For the purpose of estimating the order of magnitude of $K$ we take the value of $v_s$ at ${\bf q}_c$ to be the transverse velocity $\sim (2/3) v_L$ and the difference in velocity with the TO mode at ${\bf q}_c$ to be $1/10$ of this.  The value of ${\bf q}_c$ is taken to be $0.3 (2 \pi/a)$. $c=(c_{11}-c_{12}) \sim 10^{11} \textrm{~Nm}^{-2}$ is an elastic constant (taken from \cite{Miller}), $a=6.32 \textrm{~\AA}$ is the lattice parameter and $\rho=6400 \textrm{~Kgm}^{-3}$ is the density.  These estimates give 
$$K \sim 0.08 ~ Q^2\textrm{~[meV]}^3.$$
Dimensionless parameter $Q$ relates the strain ($\epsilon_{11}-\epsilon_{12}$) to the square of the displacement parameter $\tau$ (the displacement of the Sn relative to Te along the [111] direction divided by $\sqrt{3} a$). 
The constant $K$ may then be estimated from fitting the above formula to the data in figure 4b to be $K = 22 \textrm{~(meV)}^3$, which gives $Q \sim 17$. 

\begin{figure}[h]
\includegraphics[scale=0.6]{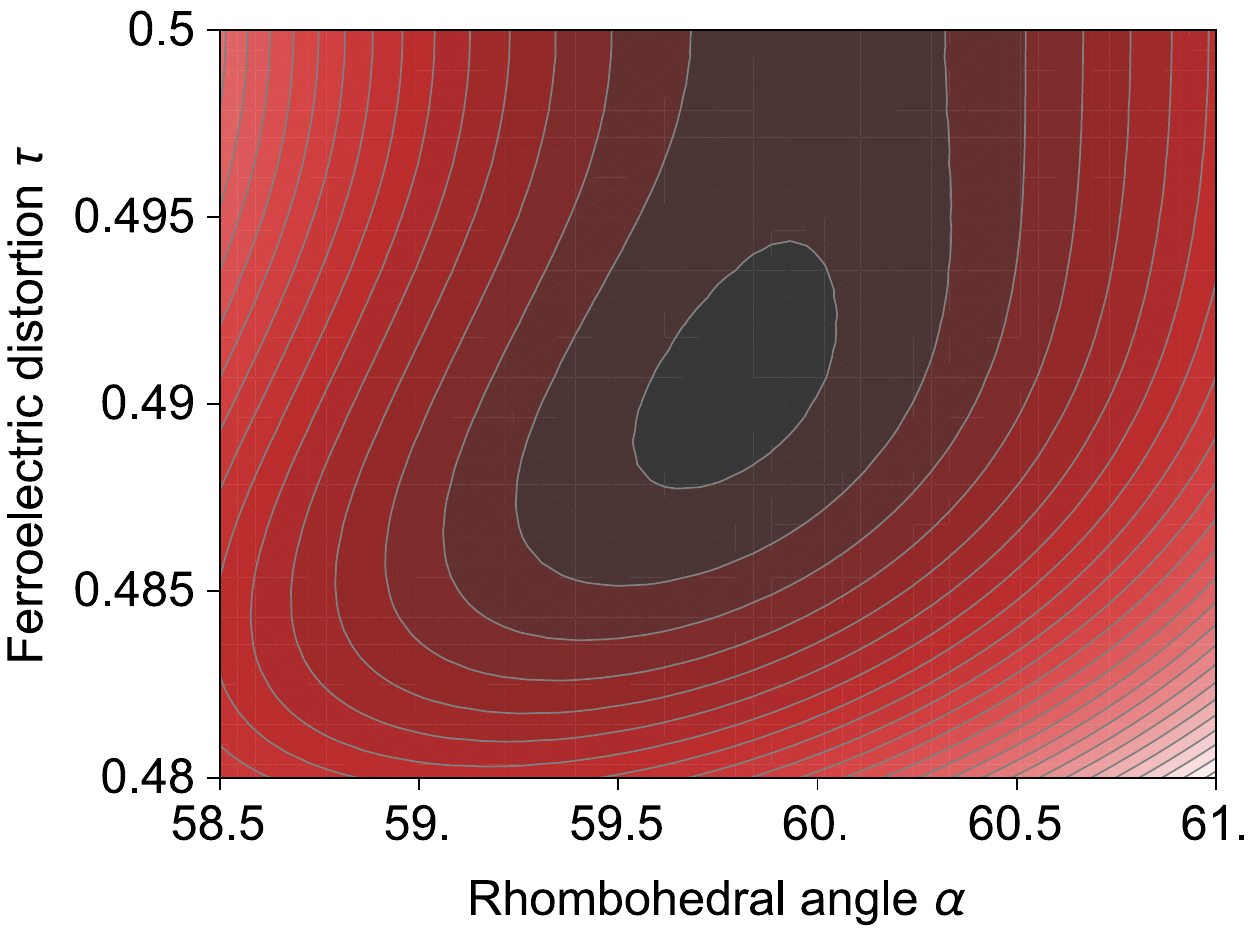}
\caption{ \label{fig}  Shows a graph of iso-energy contours as a function of the rhombohedral angle and ferroelectric displacement}
\end{figure}

Clearly the above analysis contains a large number of approximations. However the general trend that the width of the TO phonon peak increases with temperature with in addition a peak at 75 K due to the softening of the TO mode is robust.  Given the crude nature of the approximations the quality of the fit to the observed temperature dependence obtained is somewhat fortuitous. The analysis nevertheless suffices for providing an order of magnitude estimate for the required $Q$ if the line width is indeed due to this mechanism.  It is then instructive to compare this estimate for Q with that obtained from DFT.  The DFT value can be deduced from supplementary figure 7, which shows the iso-energy contours as a function of the rhombohedral distortion and ferroelectric displacement.  The value obtained at the optimum values of $\delta\alpha=0.2^{\circ}$ and $\delta\tau = 0.009$ is $Q=13.4$.  The agreement of the order of magnitude estimate based on the phonon line width, DFT and the measured $\alpha$ and $\tau $ suggest that anharmonicity may indeed account for the the phonon line width.


\end{document}